\documentstyle[12pt]{article}
\begin{document}
\title{\Large\bf{On covariant $\kappa-$symmetry fixing and the relation between the NSR string and the Type II GS superstring}}
\author{Dmitriy V. Uvarov\thanks{E-mail address: d\_uvarov@hotmail.com, uvarov@kipt.kharkov.ua}\\
\normalsize\it{NSC Kharkov Institute of Physics and Technology}\\
\normalsize\it{61108 Kharkov, Ukraine}}
\date{}
\maketitle
\begin{abstract}
By considering the superembedding equation for the Type II superstring we derive the classical relation between the NSR string and the Type II GS superstring Grassmannian variables. The connection between the actions of these two models is also established. Then introducing the proper twistor-like Lorentz harmonic variables we fix $\kappa-$symmetry of the GS formulation in the manifestly SO(1,9) Lorentz covariant manner and establish the relation between the gauge-fixed variables of the NSR and the Type II GS models.
Pacs: 11.30.P, 12.60.J
\end{abstract}
\section{Introduction}
In Refs.\cite{VZ},\cite{STVZ}  was proved the classical equivalence between the massless $N=1$ Brink-Schwarz superparticle \cite{BrS} and the massless spinning particle \cite{dh}, possessing the $n=1$ local worldsheet supersymmetry. However, their first-quantized spectra of states are different. On the other hand, it is well known that in $D=10$ the GSO-projected NSR string \cite{GSO},\cite{NSR} and the GS superstring \cite{GS},\cite{GSW} describe the same set of quantum states. Thus, there naturally arises a question of establishing a classical relation between these models in the manifestly $SO(1,9)$ Lorentz covariant manner\footnote{In \cite{GSW}  such an equivalence between the NSR string and the Type II GS superstring was proved by imposing the $SO(1,9)$ Lorentz noncovariant light-cone gauge to fix $\kappa-$symmetry.}. Although some interesting results towards a solution of this problem were obtained in \cite{Berk},\cite{APT},\cite{DP} this issue is far from being clear. 

The crucial point in solving this problem is to find a Cartan-Penrose-type relation between the NSR string and the Type II GS superstring variables $\psi^m_{\pm}$ and $\theta^{\alpha1,2}$. We suggest such a relation. It is a  direct generalization of that of Ref.\cite{VZ},\cite{STVZ} for particles and involves commumting $D=10$ MW spinors $\lambda^\alpha_{\pm}$, which are the superpartners of $\theta^{\alpha1,2}$ with respect to the $n=(1|1)$ local worldsheet supersymmetry. It is  this $n=(1|1)$ local worldsheet supersymmetry that the NSR string possesses.

Another issue which we concern in this paper is the manifestly  $SO(1,9)$ Lorentz covariant $\kappa-$symmetry fixing for the Type II GS superstring model. This allows one to deal only with the physical variables and becomes especially important when trying to simplify the GS superstring action in curved backgrounds, in particular, in the intensively studied now AdS ones \cite{MT}.  To gauge away $\kappa-$symmetry in the manifestly covariant way we use twistor-like Lorentz-harmonic variables, parametrizing the coset space $SO(1,9)/SO(1,1)\times SO(8)$ \cite{Sokatchev}-\cite{BZbrane} \footnote{The concept of harmonic variables was originally introduced in Ref.\cite{GIKOS} to describe gauge theories with extended supersymmetry.} and decompose Grassmannian spinors $\theta^{\alpha 1,2}$ on them. We also relate $\kappa-$symmetry fixed variables $\theta^-_{\dot A}$ and $\theta^+_{A}$ to the NSR string physical variables $\varphi^i_{\pm}$, which are the orthogonal to the worldsheet components of $\psi^m_{\pm}$. 

\section{Relation between the NSR string and the Type II GS superstring}

As it is known from the superembedding approach\footnote{The first models invariant under  both the local worldsheet supersymmetry and the local target-space one were proposed in \cite{Gates} and paved the way for the so-called spinning superparticle and spinning superstring models \cite{spin}. They correspond to the unresricted embedding and describe more physical states than the conventional superparticle and superstring theories.} (for review see \cite{DS}), an embedding of a supersurface into a target-superspace is governed by the so called geometrodynamical equation \cite{VZ},\cite{STVZ},\cite{Galperin}-\cite{HRS}, which  asserts that the pullback of the target-superspace supervielbein bosonic components onto the supersurface Grassmannian directions has to vanish, i.e. 
\begin{equation}\label{embed}
{\cal D}_{\hat q}Z^ME_M^a(Z)=0,
\end{equation}
where $Z^M$ is the  condensed notation for the superspace coordinates, considered as the  worldvolume scalar superfields, $E_M^a(Z)$ are the tangent space vector components of the target-superspace supervielbein 1-form and ${\cal D}_{\hat q}$ is the supersurface Grassmannian covariant derivative. Index $\hat q$ stands for the direct product of the supersurface Lorentz group $SO(1,p)$ spinor index and the one corresponding to  the fundamental representation(s) of the automorphisms group of the extended supersymmetry on the worldvolume. Written in such a form superemdedding equation is valid for all known types of branes coupled to corresponding supergravity backgrounds. We, however, will concentrate on the Type II superstrings embedded into flat $D=10$ target-superspace with the bosonic metrics $\eta^{mn}=(+,-,...,-)$. Then (\ref{embed}) reads
\begin{equation}
\Pi^m_{\pm q}={\cal D}_{\pm q}X^m-i\left({\cal D}_{\pm q}\Theta^{\alpha 1}\sigma^m_{\alpha \beta}\Theta^{\beta 1}+{\cal D}_{\pm q}\Theta_{\alpha}^2\tilde\sigma^{m\alpha \beta}\Theta_{\beta}^2\right)=0,
\end{equation}
for the Type IIA case and 
\begin{equation}
\Pi^m_{+q\atop -\dot q}={\cal D}_{+q\atop -\dot q}X^m-i\left({\cal D}_{+q\atop -\dot q}\Theta^{\alpha 1}\sigma^m_{\alpha \beta}\Theta^{\beta 1}+{\cal D}_{+q\atop -\dot q}\Theta^{\alpha 2}\sigma^m_{\alpha \beta}\Theta^{\beta 2}\right)=0,
\end{equation}
for the Type IIB case, where Grassmannian superfields $\Theta^{\alpha1}(\Theta_{\alpha}^2)(\sigma^{\pm2},\eta^{\pm q})$ (Type IIA) or $\Theta^{\alpha1,2}(\sigma^{\pm2},\eta^{+q\atop -\dot q})$ (Type IIB) are the worldsheet scalars and $D=10$ MW spinors, bosonic superfield $X^m(\sigma^{\pm2},\eta^{\pm q})$ (Type IIA) or $X^m(\sigma^{\pm2},\eta^{+q\atop -\dot q})$ (Type IIB) is also the worldsheet scalar, but $D=10$ vector. The superworldsheet is parametrized by 2 bosonic light-cone coordinates $\sigma^{\pm2}$ and $2n$ Grassmannian ones $\eta^{\pm q}$ or  $\eta^{+q\atop -\dot q}$  $(q, \dot q=1,...,n)$, that are $d=2$ MW spinors with different chiralities. So $n$, that can take values from 1 to 8, denotes the number of (anti)chiral local worlsheet supersymmetries. Here we will consider the Type IIB case as a basic one, although all the essential differences concerning the Type IIA case will be outlined.

It is known that in all classically allowed dimensions, except $D=3$, this superemedding equation yields the superstring equations of motion, thus unabling the construction of the conventional doubly supersymmetric superfield actions for the Type II superstrings in all cases but $D=3$ \cite{Galperin}, \cite{BPSTV}. Nonetheless, doubly supersymmetric superfield actions were found for the superparticles \cite{STV}-\cite{BMS} and the  heterotic superstrings \cite{Berkovits1}-\cite{BCSV} in diverse dimensions with various numbers of local worldsheet supersymmetries, as well as, for the nullstrings \cite{BSTV} and the $D=4\ \ N=1$ supermembrane \cite{HRS}.

In order to establish a relation between the NSR string and the Type II superstring we need to analyse this superembedding equation for the $n=(1|1)$ worldsheet superspace, parametrized by only two Grassmannian variables $\eta^{\pm}$, which corresponds to the NSR string type worldsheet supersymmetry. Note, that the $n=(1|1)$ worldsheet supergravity can be considered superconformally flat \cite{Howe}. Thus, component expansions of the worldsheet superfields $X^m$, $\Theta^{\alpha1,2}$ acquire the simplest form
\begin{equation}
X^m(\sigma^{\pm2},\eta^{\pm})=X^m(\sigma^{\pm2})+\frac{i}{\sqrt{8}}\eta^+\psi^m_+(\sigma^{\pm2})+\frac{i}{\sqrt{8}}\eta^-\psi^m_-(\sigma^{\pm2})+i\eta^+\eta^-F^m(\sigma^{\pm2}),
\end{equation}
\begin{equation}
\Theta^{\alpha1,2}(\sigma^{\pm2},\eta^{\pm})=\theta^{\alpha1,2}(\sigma^{\pm2})+\eta^+\lambda^{\alpha1,2}_+(\sigma^{\pm2})+\eta^-\lambda^{\alpha1,2}_-(\sigma^{\pm2})+i\eta^+\eta^-\rho^{\alpha1,2}(\sigma^{\pm2}).
\end{equation}
 $X^m$ and $\theta^{\alpha1,2}$ are ordinary GS variables, $\psi^m_{\pm}$ are the NSR Grassmannian variables, $\lambda^{\alpha1,2}_{\pm}$ are the stringy twistor-like variables, $F^m$ and $\rho^{\alpha1,2}$ are redundant auxiliary ones. Covariant derivatives then look like
\begin{equation}
{\cal D}_{\pm}=E^{-1/2}D^0_{\pm}=E^{-1/2}\left(\frac{\partial}{\partial\eta^{\pm}}-i\eta^{\pm}\partial_{\pm2}\right),
\end{equation}
where superscript ${}^0$ corresponds to the flat superworldsheet.  They satisfy the following algebra
\begin{equation}
D^0_+D^0_+=-i\partial_{+2},\ \ D^0_-D^0_-=-i\partial_{-2}, \{D^0_+,D^0_-\}=0.
\end{equation}
Fixing the superconformal gauge we impose the chirality conditions on $\Theta^{\alpha1,2}$ superfields \cite{VZ}
\begin{equation}\label{ms}
{\cal D}_-\Theta^{\alpha1}={\cal D}_+\Theta^{\alpha2}=0,
\end{equation}
which on component level are equivalent to 
\begin{equation}\label{redund}
\lambda^{\alpha1}_-=\rho^{\alpha1}=0,\ \ \lambda^{\alpha2}_+=\rho^{\alpha2}=0;
\end{equation}
\begin{equation}\label{chir}
\partial_{-2}\theta^{\alpha1}=\partial_{-2}\lambda^{\alpha1}_+\equiv\partial_{-2}\lambda^{\alpha}_+=0,\ 
\partial_{+2}\theta^{\alpha2}=\partial_{+2}\lambda^{\alpha2}_-\equiv\partial_{+2}\lambda^{\alpha}_-=0.
\end{equation}
Conditions (\ref{ms}) contain equations of motion and thus put the variables on the mass shell. 
Upon utilization of (\ref{redund}) superembedding equation yields
\begin{equation}\label{twistor}
\psi^m_-=\sqrt{8}\lambda^\alpha _-\sigma^m_{\alpha \beta}\theta^{\beta 2},\ \ \psi^m_+=\sqrt{8}\lambda^\alpha_+\sigma^m_{\alpha \beta}\theta^{\beta 1};
\end{equation}
\begin{equation}\label{Vir}
\Pi^m_{\pm2}=\partial_{\pm2}X^m-i\partial_{\pm2}\theta^1\sigma^m\theta^1-i\partial_{\pm2}\theta^2\sigma^m\theta^2=\lambda_{\pm}\sigma^m\lambda_{\pm};
\footnote{ For the Type IIA case we have $\Pi^m_{+2}=\lambda^\alpha_{+}\sigma^m_{\alpha\beta}\lambda^\beta_{+}$, $\Pi_{-2}^m=\lambda_{\alpha -}\tilde\sigma^{m\alpha\beta}\lambda_{\beta -}$.}
\end{equation}
\begin{equation}
F^m=0.
\end{equation}
Applying further equation (\ref{chir}) one recovers the NSR string fermionic equations of motion
\begin{equation}\label{NSReqm}
\partial_{+2}\psi^m_-=0,\ \partial_{-2}\psi^m_+=0.
\end{equation}

Let us discuss some properties of the obtained formulae. Reprsentation (\ref{twistor}) connects in a natural way the NSR string and the Type II GS superstring fermionic variables. The NSR string Grassmannian vectors $\psi^m_{\pm}$ contain $9+9$ components as a result of the two supercurrent constraints. However, as will be seen below, proper solution of these constraints ensures dropping out of the two extra components of $\psi^m_{\pm}$ from the NSR string action. So, actually only their $8+8$ physical components contribute to the action. On the other hand, among $16+16$ components of the two MW spinors $\theta^{\alpha 1,2}$ after explicit fixing of $\kappa-$symmetry (see Sec.3) there remain $8+8$ components. Thus, on the constraint shell there is the same number of the Grassmannian degrees of freedom in  both formulations of the string theory, as it should be. In the next  Section we will find manifest expressions for the physical variables in the NSR and the GS models and relate them to  each other.
Representation (\ref{Vir}) solves the Type II GS superstring Virasoro constraints since the vectors $\lambda_+\sigma^m\lambda_+$ and $\lambda_-\sigma^m\lambda_-$ are light-like due to  the famous $10D$ permutation relation
\begin{equation}\label{perm}
\sigma^m_{\alpha\beta}\sigma_{m\gamma\delta}+\sigma^m_{\alpha\delta}\sigma_{m\beta\gamma}+\sigma^m_{\alpha\gamma}\sigma_{m\delta\beta}=0.
\end{equation}
The NSR string and the GS superstring equations of motion are satisfied by virtue of (\ref{chir}-\ref{Vir}).

Let  us consider the NSR string constraints. The supercurrent constraints 
\begin{equation}
\psi^m_+\partial_{+2}X_m=\psi^m_-\partial_{-2}X_m=0
\end{equation}
after substitution of the representations (\ref{chir},\ref{twistor},\ref{Vir}) give rise to the following equations
\begin{equation}\label{teta}
\partial_{+2}\theta^{\alpha1}=d_+\lambda^\alpha_++e_{+2}\theta^{\alpha 1},\ \ \partial_{-2}\theta^{\alpha2}=d_-\lambda^\alpha_-+e_{-2}\theta^{\alpha2}.
\end{equation}
The fact that equations (\ref{teta}) contain worldsheet superreparametrization-like terms with only two  arbitrary functions $d_{\pm}(\sigma^{\pm2})$, as was shown in \cite{STVZ}, amounts to all but two $\kappa-$symmetry  parameteres being fixed. The rest are identified with the $n=(1|1)$ worldsheet superreparametrization transformations in order to establish the relation with the NSR string. 
The stress-tensor constraints 
\begin{equation}
\partial_{\pm2}X^m\partial_{\pm2}X_m-\frac{i}{2}\psi_{\pm}\partial_{\pm2}\psi_{\pm}=0
\end{equation}
after substitution of the representations for $\partial_{\pm2}X^m$ (\ref{Vir}) reduce to
\begin{equation}\label{stress}
\psi_{\pm}\partial_{\pm2}\psi_{\pm}=0,
\end{equation}
since the vectors $\partial_{\pm2}X^m$ are light-like as a result of the equations of motion for $\theta^{1,2}$ (\ref{chir},\ref{teta}) and the permutation formula for the $D=10$ $\sigma-$matrices (\ref{perm}).
From the constraints (\ref{stress}) there follow the equations of motion for the commuting spinors $\lambda^\alpha_{\pm}$
\begin{equation}\label{dlambda}
\partial_{+2}\lambda^\alpha_+=f_{+2}\lambda^\alpha _++g_{+3}\theta^{\alpha1},\ \ \partial_{-2}\lambda^\alpha_-=f_{-2}\lambda^\alpha_-+g_{-3}\theta^{\alpha2}.
\end{equation}
Equations (\ref{teta},\ref{dlambda}) lead to the following expressions for $\partial_{\pm2}\psi^m_\pm$:
\begin{equation}\label{dpsi}
\partial_{+2}\psi^m_+=(e_{+2}+f_{+2})\psi^m_++d_+\lambda_+\sigma^m\lambda_+, \ \partial_{-2}\psi^m_-=(e_{-2}+f_{-2})\psi^m_-+d_-\lambda_-\sigma^m\lambda_-.
\end{equation}

After taking into account (\ref{twistor}, \ref{Vir}) it is  possible to establish the connection  between the NSR and the Type II GS string actions
\begin{equation}
S_{GS}=S_{NSR}+\Delta S,
\end{equation}
where 
\begin{equation}
S_{NSR}=-\frac{2}{c\alpha^\prime}\partial_{+2}X^m\partial_{-2}X_m+\frac{i}{c\alpha^\prime}\left(\psi^m_+\partial_{-2}\psi_{m+}+\psi^m_-\partial_{+2}\psi_{m-}\right),
\end{equation}
\begin{equation}\begin{array}{c}
\Delta S=-\frac{4}{c\alpha^\prime}\partial_{-2}\theta^1\sigma^m\theta^1\partial_{+2}\theta^2\sigma_m\theta^2-
\frac{2}{c\alpha^\prime}\left(\partial_{+2}\theta^1\sigma^m\theta^1\partial_{-2}\theta^1\sigma_m\theta^1+\partial_{+2}\theta^2\sigma^m\theta^2\partial_{-2}\theta^2\sigma_m\theta^2\right)\\[0.4cm]
-\frac{8i}{c\alpha^\prime}\left(\lambda_+\sigma^m\theta^1\partial_{-2}\lambda_+\sigma_m\theta ^1+\lambda_-\sigma^m\theta^2\partial_{+2}\lambda_-\sigma_m\theta ^2\right).
\end{array}\end{equation}
$\Delta S$ vanishes on the mass shell (\ref{chir}).

\section{Manifestly $SO(1,9)$ Lorentz covariant $\kappa-$symmetry fixing}

Now let us concern the issue of $\kappa-$symmetry fixing for the Type II GS superstring. To this end let us consider the classically equivalent twistor-like Lorentz harmonic formulation for the Type IIB GS superstring \cite{BZstring}:
\begin{equation}\label{gs}
\begin{array}{c}
{\displaystyle
S=\int e\left(-(4\alpha^\prime)^{-1/2}\left(e^{\mu[+2]}u^{[-2]}_m+e^{\mu[-2]}u^{[+2]}_m\right)\omega^m_\mu+c\right)-}\\
{\displaystyle
\frac{1}{c\alpha^\prime}\int \epsilon^{\mu\nu}\left[i\omega^m_\mu\left(\partial_\nu\theta^1\sigma_m\theta^1-\partial_\nu\theta^2\sigma_m\theta^2\right)+\partial_\mu\theta^1\sigma_m\theta^1\partial_\nu\theta^2\sigma_m\theta^2\right]}.\end{array}
\footnote{It is this action, that an  unknown doubly supersymmetric superfield one should reduce to after integrating out the superworldsheet Grassmannian variables and elimination of auxiliary ones.}
\end{equation}
 In the Type IIA case one should replace $\partial_\nu\theta^{\alpha 2}\sigma_{m\alpha \beta }\theta^{\beta 2}$ with $\partial_\nu\theta^2_\alpha\tilde \sigma_m^{\alpha \beta }\theta^2_\beta $. In addition to the variables which are present in the standart GS superstring formulation \cite{GS} it contains 
the worldsheet zweinbein $e^{\mu[\pm2]}$ and the light-like Lorentz frame vectors $u^{m[\pm2]}$ tangent to the string worldsheet. These light-like Lorentz frame vectors together with the orthogonal to the worldsheet  ones $u^{m(i)}$ $((i)=1,...,8)$ constitute a complete orthonormal basis 
one can use to expand any  $D=10$ Minkowski vector. Lorentz frame vectors can be presented as the bilinear combinations of the spinor harmonics $v^a_\alpha=(v^{-}_{\alpha\dot A},v^{+}_{\alpha A})$ or their inverse $(v^{-1})^\alpha_a=(v^{\alpha-}_A,v^{\alpha+}_{\dot A})$ ($(v^{-1})^\alpha_a v^b_\alpha=\delta^b_a$):
\begin{equation}
u^{[+2]}_m=\frac18(v^{+}_{\alpha A}\tilde \sigma_m^{\alpha\beta}v^{+}_{\beta A})=\frac18(v^{\alpha+}_{\dot A}
\sigma_{m\alpha\beta}v^{\beta+}_{\dot A}),
\end{equation}
\begin{equation}
u^{[-2]}_m=\frac18(v^{-}_{\alpha\dot A}\tilde \sigma_m^{\alpha\beta}v^{-}_{\beta\dot A})=\frac18(v^{\alpha-}_A\sigma_{m\alpha\beta}v^{\beta-}_A),
\end{equation}
\begin{equation}
u^{(i)}_m=\frac18(v^{+}_{\alpha A}\tilde\sigma_m^{\alpha\beta}v^{-}_{\beta\dot A})\gamma^i_{A\dot A}=-\frac18(v^{\alpha-}_A\sigma_{m\alpha\beta}v^{\beta+}_{\dot A})\gamma^i_{A\dot A}.
\end{equation}
They are orthonormal
\begin{equation}\label{harmonics}
u^{[\mp2]}\cdot u^{[\pm2]}=0, u^{[+2]}\cdot u^{[-2]}=2,\\
u^{[\pm2]}\cdot u^{(i)}=0, u^{(i)}\cdot u^{(j)}=-\delta^{(i)(j)}
\end{equation}
as a result of certain harmonicity conditions imposed on the spinor harmonics that reduce the number of the independent variables in the spinor harmonics to the dimension of the $SO(1,9)$ Lorentz group equal to 45 \cite{harmonics}. Lorentz frame vector harmonics satisfy the following differential equations
\begin{equation}\label{derive}\begin{array}{c}
\partial_{\pm2} u^{[+2]}_m=\Omega^{(0)}_{\pm2} u^{[+2]}_m+\Omega^{[+2](i)}_{\pm2} u^{(i)}_m,\\[0.3cm]
\partial_{\pm2} u^{[-2]}_m=-\Omega^{(0)}_{\pm2} u^{[-2]}_m+\Omega^{[-2](i)}_{\pm2} u^{(i)}_m,\\[0.3cm]
\partial_{\pm2} u^{(i)}_{m}=\frac12\Omega^{[+2](i)}_{\pm2} u_m^{[-2]}+
\frac12\Omega^{[-2](i)}_{\pm2} u_m^{[+2]}+\Omega^{(i)(j)}_{\pm2} u^{(j)}_m.
\end{array}\end{equation}
These are the only possible equations compatible with the orthonormality conditions (\ref{harmonics}).  Coefficients $\Omega$ in (\ref{derive}) are the $SO(1,1)\times SO(8)$ decomposed $SO(1,9)$ Cartan forms. From the embedding theory point of view \cite{Eis},\cite{Barbashov},\cite{Lund},\cite{Omnes},\cite{Zhelt},\cite{BIKU} $\Omega^{(i)(j)}_{\pm2}$ can be identified with the torsion (third fundamental form) components, $\Omega^{[+2](i)}_{\pm2}$ and $\Omega^{[-2](i)}_{\mp2}$ with the second fundamental form components and $\Omega^{(0)}_{\pm2}$ with the 2d spin connection. Integrability conditions of Eqs.(\ref{derive}) are Gauss, Peterson-Kodacci and Ricci equations \cite{Eis},\cite{Barbashov}. 

Action (\ref{gs}) is invariant under the following $\kappa-$symmetry gauge transformations with the local parameters $\kappa^+_A$ and $\kappa^-_{\dot A}$:
\begin{equation}\label{kappa}
\begin{array}{c}
\delta\theta^{\alpha 1}=v^{\alpha-}_A\kappa^+_A,\ \ \delta\theta^{\alpha 2}=v^{\alpha+}_{\dot A}\kappa^-_{\dot A},\\[0.3cm]
\delta X^m=i\left(\kappa^+_A v^{\alpha-}_A\sigma^m_{\alpha\beta}\theta^{\beta 1}+\kappa^-_{\dot A}v^{\alpha+}_{\dot A}
\sigma^m_{\alpha\beta}\theta^{\beta 2}\right),\\[0.4cm]
\delta\left(ee^{\mu[+2]}\right)=\frac{4i}{c(\alpha\prime)^{1/2}}\kappa^+_A\varepsilon^{\mu\nu}\partial_\nu \theta^{\alpha 1}v^{+}_{\alpha A},\ \ \delta\left(ee^{\mu[-2]}\right)=-\frac{4i}{c(\alpha \prime)^{1/2}}\kappa^-_{\dot A}\varepsilon^{\mu\nu}\partial_\nu \theta^{\alpha 2}v^{-}_{\alpha\dot A},\\[0.4cm]
\delta u^{[+2]}_m=\frac{2i}{c(\alpha\prime)^{1/2}}e^{\mu[+2]}w^{(i)}_\mu u^{(i)}_m,\ \ \delta u^{[-2]}_m=-\frac{2i}{c(\alpha\prime)^{1/2}}e^{\mu[-2]}w^{(i)}_\mu u^{(i)}_m,\\
\end{array}\end{equation}
where $w^{(i)}_\mu=\left(\kappa^+_B\gamma^i_{B\dot B}\partial_\mu \theta^{\alpha 1}v^{-}_{\alpha\dot B}-\kappa^-_{\dot B}\tilde\gamma^i_{\dot B B}\partial_\mu \theta^{\alpha 2}v^{+}_{\alpha B}\right)$
\footnote{For  the Type IIA case $\kappa-$symmetry transformations read:
\begin{equation}\label{kappaA}
\begin{array}{c}
\delta\theta^{\alpha 1}=v^{\alpha-}_A\kappa^+_A,\ \ \delta\theta_\alpha^2=v^+_{\alpha A}\kappa^-_{ A},\\[0.3cm]
\delta X^m=i\left(\kappa^+_A v^{\alpha-}_A\sigma^m_{\alpha\beta}\theta^{\beta 1}+\kappa^-_{A}v^+_{\alpha A}
\tilde\sigma^{m\alpha\beta}\theta_\beta^2\right),\\[0.4cm]
\delta\left(ee^{\mu[+2]}\right)=\frac{4i}{c(\alpha^\prime)^{1/2}}\kappa^+_A\varepsilon^{\mu\nu}\partial_\nu \theta^{\alpha 1}v^{+}_{\alpha A},\ \ \delta\left(ee^{\mu[-2]}\right)=-\frac{4i}{c(\alpha^\prime)^{1/2}}\kappa^-_{A}\varepsilon^{\mu\nu}\partial_\nu \theta_\alpha^2v^{\alpha-}_{A},\\[0.4cm]
\delta u^{[+2]}_m=-\frac{2i}{c(\alpha^\prime)^{1/2}}e^{\mu[+2]}w^{(i)}_\mu u^{(i)}_m,\ \ \delta u^{[-2]}_m=\frac{2i}{c(\alpha^\prime)^{1/2}}e^{\mu[-2]}w^{(i)}_\mu u^{(i)}_m,\\
\end{array}\end{equation}
where $w^{(i)}_\mu=-\left(\kappa^+_B\gamma^i_{B\dot B}\partial_\mu \theta^{\alpha 1}v^{-}_{\alpha\dot B}+\kappa^-_{B}\gamma^i_{B\dot B}\partial_\mu \theta_\alpha^2v^{\alpha+}_{\dot B}\right)$.}.

To fix $\kappa-$symmetry gauge it is  useful to expand Grassmannian variables using the spinor harmonics:
\begin{equation}\label{newteta}
\theta^{\alpha 1(2)}=v^{\alpha-}_A \theta^{1(2)+}_A+v^{\alpha+}_{\dot A}\theta^{1(2)-}_{\dot A}.
\end{equation}
$\kappa-$Symmetry transformations for the introduced  variables look as follows:
\begin{equation}\label{teta1}
\delta\theta^{1-}_{\dot A}=-\frac{i}{c(\alpha^\prime)^{1/2}}\theta^{1+}_A\gamma^i_{A\dot A}e^{\mu[-2]}w^{(i)}_\mu,\ \delta\theta^{1+}_A=\kappa^+_A+\frac{i}{c(\alpha^\prime)^{1/2}}\theta^{1-}_{\dot A}\tilde\gamma^i_{\dot A A}e^{\mu[+2]}w^{(i)}_\mu,
\end{equation}
\begin{equation}\label{teta2}
\delta\theta^{2+}_A=\frac{i}{c(\alpha^\prime)^{1/2}}\theta^{2-}_{\dot A}\tilde\gamma^i_{\dot A A}e^{\mu[+2]}w^{(i)}_\mu,\ \delta\theta^{2-}_{\dot A}=\kappa^-_{\dot A}-\frac{i}{c(\alpha^\prime)^{1/2}}\theta^{2+}_{A}\gamma^i_{A\dot A}e^{\mu[-2]}w^{(i)}_\mu\footnote{ Note that  for the null-string terms proportional  to $(\alpha^{\prime})^{-1}$ vanish  since $\alpha^\prime\rightarrow\infty$ \cite{FortP}.}.
\end{equation}
$\theta^{1+}_A$ and $\theta^{2-}_{\dot A}$ are pure gauge variables, so we are able to impose the following $\kappa-$symmetry fixing conditions:
\begin{equation}\label{kappafix}
\theta^{1+}_A=0,\ \ \theta^{2-}_{\dot A}=0.\footnote{In the Type IIA case $\kappa-$symmetry fixing conditions look like $\theta^{1+}_A=0, \theta^{2-}_{A}=0$.}
\end{equation}
In this gauge the remaining variables $\theta^{1-}_{\dot A}\equiv\theta^-_{\dot A}$ and $\theta^{2+}_A\equiv\theta^+_A$ are $\kappa-$invariant as is seen from (\ref{teta1}),(\ref{teta2}). Note, that they are the worldsheet MW spinors, whereas original variables $\theta^{\alpha1,2}$ were the worldsheet scalars.
$\kappa-$Symmetry fixed action (\ref{gs}) written in these new variables acquires the form
\begin{equation}\label{gsfixed}\begin{array}{c}
\displaystyle{S_{fixed}=\int e\left[-(\!\alpha^\prime)^{-1/2}e^\mu_{[-2]}\left(\!D_{\mu}x^{[-2]}\!-\!2i\tilde D_{\mu }\theta^-\!\cdot\!\theta^-\!\right)\!-\!(\!\alpha^\prime)^{-1/2}e^\mu_{[+2]}\left(D_{\mu}x^{[+2]}\!-\!2iD_{\mu}\theta^+\!\cdot\!\theta^+\!\right)\!+\!c\right]}\\[0.4cm]
\displaystyle{\!\!-\frac{i}{c\alpha^\prime}\int\!\!\epsilon^{\mu \nu}\!\left[\!\left(\!D_{\mu}x^{[+2]}\!-\!2iD_{\mu}\theta^+\!\!\cdot\!\theta^+\!\right)\!\tilde D_{\nu}\!\theta^-\!\!\cdot\!\theta^-\!\!-\!\!\left(\!D_{\mu}x^{[-2]}\!-\!2i\tilde D_{\mu}\theta^-\!\!\cdot\!\theta^-\!\right)\!D_{\nu}\theta^+\!\!\cdot\!\theta^+\!\!-\!\!2i\tilde D_{\mu}\theta^-\!\!\cdot\!\theta^-\!D_{\nu}\theta^+\!\!\cdot\!\theta^+\right.}\\[0.4cm]
\displaystyle{\left.
-\frac12\!\left(\!D_{\mu}x^{(i)}\!\!-\!\!\frac14\!\left(\!\theta^-\tilde\gamma^{ij}\theta^-\!\right)\!\!\Omega^{[+2](j)}_{\mu}\!\!-\!\!\frac14\!\left(\!\theta^+\gamma^{ij}\theta^+\!\right)\!\Omega^{[-2](j)}_{\mu}\!\right)\!\!\left(\!\left(\!\theta^-\tilde\gamma^{ij}\theta^-\!\right)\!\Omega^{[+2](j)}_{\nu}\!-\!\left(\!\theta^+\gamma^{ij}\theta^+\!\right)\!\Omega^{[-2](j)}_{\nu}\!\right)\!\right].}
\end{array}\end{equation}
We expanded bosonic coordinates $x^m$ on the vector harmonics 
\begin{equation}\label{x}
x^m=u^{m(n)}x_{(n)}\equiv\frac12 u^{m[+2]}x^{[-2]}+\frac12 u^{m[-2]}x^{[+2]}-u^{(i)}_mx^{(i)}
\end{equation}
in order to get rid of harmonics in the action and also introduced vector covariant derivatives $D_\mu x^{(n)}=\partial_\mu x^{(n)}+\Omega_{\mu}^{(n)(l)} x_{(l)},$ whose decompositions read
\begin{eqnarray}
D_\mu x^{[\pm2]}=\partial_\mu x^{[\pm2]}\mp\Omega^{(0)}_\mu x^{[\pm2]}-\Omega^{[\pm2](i)}_\mu x^{(i)},\\
D_\mu x^{(i)}=\partial_\mu x^{(i)}-\Omega^{(i)(j)}_\mu x^j-\frac12\Omega^{[+2](i)}_\mu x^{[-2]}-\frac12\Omega^{[-2](i)}_\mu x^{[+2]},
\end{eqnarray}
and the spinor ones
\begin{eqnarray}
D_{\mu}\theta^+_A=\partial_{\mu}\theta^+_A-\frac12\Omega^{(0)}_\mu\theta^+_A-\frac14\Omega^{(i)(j)}_{\mu}\gamma^{ij}_{AB}\theta^+_B,\\
\tilde D_{\mu}\theta^-_{\dot A}=\partial_{\mu}\theta^-_{\dot A}+\frac12\Omega^{(0)}_\mu\theta^-_{\dot A}-\frac14\Omega^{(i)(j)}_{\mu}\tilde\gamma^{ij}_{\dot A\dot B}\theta^-_{\dot B}.
\end{eqnarray}
Action (\ref{gsfixed}) contains the following light-cone-like terms quadratic in $\theta^{\pm}$
\begin{equation}\label{gslightcone}
S_{l.c.}\!=\!\frac{2i}{(\alpha^\prime){}^{1/2}}\!\int\! e\!\!\left[\!\left(\!1\!+\!\frac{2}{c(\alpha^\prime){}^{1/2}}\!D_{[+2]}x^{[+2]}\!\right)\!\!
\tilde D_{[-2]}\!\theta^-\!\cdot\!\theta^-\!+\!\left(\!1\!+\!\frac{2}{c(\alpha^\prime){}^{1/2}}\!D_{[-2]}x^{[-2]}\!\right)\!\!D_{[+2]}\!\theta^+\!\cdot\!\theta^+\!\right]
\end{equation}
Note, that the form of the action (\ref{gsfixed}) resembles that of Ref.\cite{Sokatchev} for superparticles.

Let us turn to the problem of establishing a connection between the Lorentz harmonic variables and the commuting spinors $\lambda^\alpha_{\pm}$ of the previous section.  For this note that the variation of the action (\ref{gs}) with respect to  the zweinbeins and harmonics produces the superstring embedding equation
\begin{equation}
\Pi^m_\mu=\frac{c(\alpha^\prime){}^{1/2}}{2}\left(e^{[-2]}_\mu u^{m[+2]}+e^{[+2]}_\mu u^{m[-2]}\right).
\end{equation}
It can be simplified by choosing the conformal gauge for the zweinbein $e^f_\mu=e^{-\phi}\delta^f_\mu$:
\begin{equation}\label{gsembed}
\Pi^m_{\pm2}={\textstyle\frac{c(\alpha\prime)^{1/2}}{2}}e^{-\phi}u^{m[\mp2]}.
\end{equation}
Equation (\ref{gsembed}) coincides with (\ref{Vir}) only if
\begin{equation}\label{coin}
\lambda_{\pm}\sigma^m\lambda_{\pm}=\frac{c(\alpha^\prime){}^{1/2}}{2}e^{-\phi}u^{m[\mp2]},
\end{equation}
thus establishing the bridge with the discussion of the preceding section. To analyse the consequences of (\ref{coin}) let us expand $\lambda^\alpha_{\pm}$ on the spinor harmonics
\begin{equation}\label{newlambda}
\lambda^\alpha _+=v^{\alpha -}_A\lambda_A+v^{\alpha+}_{\dot A}\lambda_{+2\dot A},\ \ \lambda^\alpha _-=v^{\alpha -}_A\lambda_{-2A}+v^{\alpha+}_{\dot A}\lambda_{\dot A}.
\end{equation}
Then one finds that $\lambda_{+2\dot A}=\lambda_{-2A}=0$ and $\lambda_A{}^2=\frac{c(\alpha\prime)^{1/2}}{2}e^{-\phi}, \lambda_{\dot A}{}^2=\frac{c(\alpha\prime)^{1/2}}{2}e^{-\phi}$. 
 The integrability conditions of equations (\ref{gsembed}) after using (\ref{chir}) lead to the following expressions for the $2d$ spin connection
\begin{equation}\label{omega0}
\Omega^{(0)}_{+2}=\partial_{+2}\phi, \ \ \Omega^{(0)}_{-2}=-\partial_{-2}\phi ,
\end{equation}
and the minimality conditions for the components of the second fundamental form
\begin{equation}\label{minimal}
\Omega^{[-2](i)}_{-2}=\Omega^{[+2](i)}_{+2}=0.
\end{equation}

The connection between  $\kappa-$symmetry fixed GS variables and the NSR variables can be established upon the substitution of 
\begin{equation}\label{rep}
\theta^{\alpha 1}=v^{\alpha +}_{\dot A}\theta^-_{\dot A},\ \theta^{\alpha 2}=v^{\alpha -}_A\theta^+_A,\ \lambda^\alpha_+=v^{\alpha -}_A\lambda_A,\ \lambda^\alpha _-=v^{\alpha +}_{\dot A}\lambda_{\dot A}
\end{equation}
into (\ref{twistor}) and the expansion of $\psi^m_{\pm}$ on the vector harmonics
\begin{equation}\label{psi}
\psi^m_{\pm}=u^{m(n)}\varphi_{\pm (n)}=\frac12 u^{m[+2]}\varphi^{[-2]}_{\pm}+\frac12 u^{m[-2]}\varphi^{[+2]}_{\pm}-u^{(i)}_m\varphi^{(i)}_{\pm}.
\end{equation}
As a result we obtain
\begin{equation}\label{fix}
\varphi^{[+2]}_\pm=\varphi^{[-2]}_\pm=0,\ \ \varphi^i_+=-\sqrt{8}\lambda_A\gamma ^i_{A\dot B}\theta^-_{\dot B},\ \varphi^i_-=-\sqrt{8}\lambda_{\dot A}\tilde \gamma ^i_{\dot A B}\theta^+_B.\footnote{ For the Type IIA case we have: $\theta^{\alpha 1}=v^{\alpha +}_{\dot A}\theta^-_{\dot A}, \theta_{\alpha}^2=v^-_{\alpha\dot A}\theta^+_{\dot A}, \lambda^\alpha_+=v^{\alpha -}_A\lambda^1_A, \lambda_{\alpha -}=v^+_{\alpha A}\lambda^2_{A}$, so $\varphi^i_+=-\sqrt{8}\lambda^1_A\gamma ^i_{A\dot B}\theta^-_{\dot B},\ \varphi^i_-=-\sqrt{8}\lambda^2_{A}\gamma^i_{A\dot B}\theta^+_{\dot B}$.}
\end{equation} 

The equations of motion for $\kappa-$symmetry fixed Grassmannian variables $\theta^-_{\dot A}$ and $\theta ^+_A$ and the commuting spinors $\lambda_A$ and $\lambda_{\dot A}$  one  obtaines after the substitution of (\ref{rep}) into (\ref{chir},\ref{teta},\ref{dlambda}) and taking into account the equations of motion for the spinor harmonics following from (\ref{derive})
\begin{equation}\label{tetaeq}
D_{+2}\theta^+_A=0,\qquad D_{-2}\theta^+_A=e_{-2}\theta^+_A,\qquad 
\tilde D_{+2}\theta^-_{\dot A}=e_{+2}\theta^-_{\dot A},\qquad 
\tilde D_{-2}\theta^-_{\dot A}=0.
\end{equation}
Note, that these equations can be considered from the $2d$ field theory point of view as the Dirac equations for spinors interacting with the $SO(1,1)\times SO(8)$ Yang-Mills connection. 
There also appear the following  equations for the nonzero components of the second fundamental form
\begin{equation}\label{const}
\Omega^{[-2](i)}_{+2}\tilde\gamma^i_{\dot BA}\theta^+_A=0,\qquad\Omega^{[+2](i)}_{-2}\gamma^i_{A\dot B}\theta^-_{\dot B}=0
\end{equation}
and for the coefficients in (\ref{teta})
\begin{equation}
d_+=d_-=0.
\end{equation}
As was noted before, coefficients $d_+$ and $d_-$ correspond to the two unfixed $\kappa-$symmetry trnasformations which, when establishing the relation with the NSR string, are identified with the $n=(1|1)$ worldsheet superreparametrizations.  Their nullification  signifies that we have entirely fixed $\kappa-$symmetry.

Analogously equations for $\lambda_A$ and $\lambda_{\dot A}$ are
\begin{equation}\label{l}
D_{+2}\lambda_A=f_{+2}\lambda_A,\qquad D_{-2}\lambda_{A}=0,\qquad 
\tilde D_{-2}\lambda_{\dot A}=f_{-2}\lambda_{\dot A},\qquad D_{+2}\lambda_{\dot A}=0.
\end{equation}
There also appear new equations for the components of the second fundamental form
\begin{equation}\label{newconst}
\Omega^{[-2](i)}_{+2}\tilde\gamma^i_{\dot BA}\lambda_A=g_{+3}\theta ^-_{\dot B},\ \ \Omega^{[+2](i)}_{-2}\gamma^i_{A\dot B}\lambda_{\dot B}=g_{-3}\theta^+_A.
\end{equation}

To solve equations (\ref{const}) we suggest that 
\begin{equation}
\Omega^{[-2](i)}_{+2}=\tilde \Omega^{[-10](i)}_{+2}(\theta^+)^8,\ \ \Omega^{[+2](i)}_{-2}=\tilde \Omega^{[+10](i)}_{-2}(\theta^-)^8,
\end{equation}
where $(\theta^{-})^8\equiv\varepsilon_{\dot A_1...\dot A_8}\theta^-_{\dot A_1}...\theta^-_{\dot A_8}$ and $(\theta^{+})^8\equiv\varepsilon_{A_1...A_8}\theta^+_{A_1}...\theta^+_{A_8}$.
Further, the following representation
\begin{equation}
\tilde\Omega^{[-10](i)}_{+2}=(\theta^-\tilde\gamma^{ij}\theta^-)q^{j}_{+10},\ \ \tilde\Omega^{[+10](i)}_{-2}=(\theta^+\gamma^{ij}\theta^+)q^{j}_{-10}
\end{equation}
allows to define $g_{\pm3}=8\varphi^i_{\pm}q^{i}_{\pm10}(\theta^{\pm})^8$ and to reformulate (\ref{newconst}) as 
\begin{equation}\label{sys}
(\theta^-\tilde\gamma^{ij}\theta^-)q^{k}_{+10}\gamma^{ijk}_{\dot BA}\lambda_A=0,\ \ (\theta^+\gamma^{ij}\theta^+)q^{k}_{-10}\gamma^{ijk}_{B\dot A}\lambda_{\dot A}=0.
\end{equation}
The system (\ref{sys}) has the rank equal to 4 so it admits nontrivial solutions.

The integrability conditions for  equations (\ref{tetaeq},\ref{l}) look like
\begin{eqnarray}
\partial_{+2}\Omega^{(i)(j)}_{-2}-\partial_{-2}\Omega^{(i)(j)}_{+2}+\Omega^{(k)(i)}_{+2}\Omega^{(k)(j)}_{-2}-\Omega^{(k)(i)}_{-2}\Omega^{(k)(j)}_{+2}=0,\label{om}\\
\partial_{\pm2}e_{\mp2}=\partial_{+2}\partial_{-2}\phi\label{f1}\\ 
\partial_{\pm2}f_{\mp2}=-\partial_{+2}\partial_{-2}\phi.\label{f2} 
\end{eqnarray}
Equation (\ref{om}) coincides with Ricci equation after taking into account  introduced representations for the second fundamental form components. Then Gauss equation reduces to $\partial_{+2}\partial _{-2}\phi=0$. Thus, $e_{\pm2}=e_{\pm2}(\sigma^{\pm2})$ is an arbitrary function and $f_{\pm2}=-\partial_{\pm2}\phi$ as follows from the normalization  conditions for $\lambda_A$, $\lambda_{\dot A}$.

The general solutions to equations (\ref{tetaeq},\ref{l}) read:
\begin{eqnarray}
\hspace*{-1cm}\theta^+_A=\exp\!\left(-\phi/2+\!\!\int\!\! e_{-2}(\sigma^{-2})d\sigma^{-2}\right)\!\alpha^{-1}_{AB}\theta^+_{0B},\ \theta ^-_{\dot A}=\exp\!\left(-\phi/2+\!\!\int\!\! e_{+2}(\sigma^{+2})d\sigma^{+2}\right)\!\beta^{-1}_{\dot A\dot B}\theta ^-_{0\dot B},\\
\lambda_A=\exp\left(-\phi/2\right)\alpha^{-1}_{AB}\lambda_{0B},\ \ \lambda_{\dot A}=\exp\left(-\phi/2\right)\beta^{-1}_{\dot A\dot B}\lambda_{0\dot B},
\end{eqnarray}
where the nondegenerate matrices $\alpha_{AB}$ and $\beta_{\dot A\dot B}$ satisfy the following equations
\begin{equation}\label{ab}
\partial_{\pm2}\alpha_{AB}=-\frac14\alpha_{AC}\Omega^{(i)(j)}_{\pm2}\gamma^{(i)(j)}_{CB},\ \
\partial_{\pm2}\beta_{\dot A\dot B}=-\frac14\beta_{\dot A\dot C}\Omega^{(i)(j)}_{\pm2}\tilde \gamma^{(i)(j)}_{\dot C\dot B}.
\end{equation}
The integrability conditions for these equations coincide with (\ref{om}).

Equations of motion for the NSR string physical variables $\varphi^{i}_{\pm}$ can be obtained either by differentiating (\ref{fix}) and substituting the equations of motion for $\theta^+_A$, $\theta^-_{\dot A}$, $\lambda_{A}$, $\lambda_{\dot A}$ or by substitution of (\ref{fix}) into (\ref{NSReqm},\ref{dpsi})
\begin{eqnarray}
\partial_{+2}\varphi^{(i)}_+-\Omega^{(i)(j)}_{+2}\varphi^{(j)}_+=\tilde e_{+2}\varphi^{(i)}_+,\ \ 
\partial_{-2}\varphi^{(i)}_+-\Omega^{(i)(j)}_{-2}\varphi^{(i)}_+=0,\\
\partial_{-2}\varphi^{(i)}_--\Omega^{(i)(j)}_{-2}\varphi^{(j)}_-=\tilde e_{-2}\varphi^{(i)}_-,\ \ 
\partial_{+2}\varphi^{(i)}_--\Omega^{(i)(j)}_{+2}\varphi^{(i)}_-=0,
\end{eqnarray}
where $\tilde e_{\pm2}(\sigma^{\pm2})=e_{\pm2}-\partial_{\pm2}\phi.$
Their solutions read
\begin{equation}
\varphi^i_+=\exp\left(\int\tilde e_{+2}(\sigma^{+2})d\sigma^{+2}\right)A^{-1ij}\varphi^j_{0+}, \varphi^i_-=\exp\left(\int\tilde e_{-2}(\sigma^{-2})d\sigma^{-2}\right)A^{-1ij}\varphi^j_{0-},
\end{equation}
where invertible matrix $A^{ij}$ satisfies the following system
\begin{equation}\label{A}
\partial_{-2}A^{(i)(j)}=-A^{(i)(k)}\Omega^{(k)(j)}_{-2},\ \ \partial_{+2}A^{(i)(j)}=-A^{(i)(k)}\Omega^{(k)(j)}_{+2}.
\end{equation}

After substitution of $\varphi^{i}_{\pm}$ into the NSR string action its fermionic part acquires the form
\begin{equation}\label{NSRfi}
S^{ferm}_{NSR}=-\frac{i}{c\alpha^\prime}\left(\varphi^i_+\partial_{-2}\varphi^i_+-\varphi^i_+\Omega^{ij}_{-2}\varphi^j_+\right)-
\frac{i}{c\alpha^\prime}\left(\varphi^i_-\partial_{+2}\varphi^i_--\varphi^i_-\Omega^{ij}_{+2}\varphi^j_-\right).
\end{equation}
Taking into account the fact that (\ref{fix}) contains $8d$ $\sigma-$matrices and performing corresponding Fierz rearrengements one obtains another form of the action (\ref{NSRfi})
\begin{equation}
\begin{array}{c}
S^{fermmod}_{NSR}=-\frac{32i}{{\alpha^\prime}^{1/2}}e^{-\phi}\left(\theta^+_AD_{+2}\theta^+_A+\theta^-_{\dot A}\tilde D_{-2}\theta^-_{\dot A}\right)\label{lightcone}\\[0.4cm]
-\frac{i}{3!c\alpha^\prime}\sum\limits_{\pm}\left[\psi^{ijk}_{\pm}\left(\partial_{\mp2}\psi^{ijk}_{\pm}-\Omega^{im}_{\mp2}\psi^{mjk}_{\pm}-\Omega^{jm}_{\mp2}\psi^{imk}_{\pm}-\Omega^{km}_{\mp2}\psi^{ijm}_{\pm}\right)\right],\\
\end{array}\end{equation}
where the first part is the covariatized version of the light-cone action\footnote{ It coincides with (\ref{gslightcone}), written in the conformal gauge, provided that the following equations are satisfied $D_{+2}x^{[+2]}=D_{-2}x^{[-2]}=17c(\alpha^\prime){}^{1/2}e^{-\phi}$.} \cite{GSW}, the second part involves new variables, absent in the original formulations of  both the NSR string and the GS superstring theories, namely, the Grassmannian 3-forms $\psi^{ijk}_{+}=\sqrt{8}\lambda_A\gamma^{ijk}_{A\dot A}\theta^-_{\dot A}$ and  $\psi^{ijk}_{-}=\sqrt{8}\lambda_{\dot A}\tilde\gamma^{ijk}_{\dot AA}\theta^{+}_A$. 

\section{Conclusions}
 We have considered the $n=(1|1)$ superembedding equation for the Type II superstring. It was shown to contain the relation (\ref{twistor}) between the NSR string and the Type II GS superstring variables, as well as the solution to the super-Virasoro constraints. Upon manifestly $SO(1,9)$ Lorentz covariant fixation of $\kappa-$symmetry  using the twistor-like Lorentz harmonic variables, which amounts to covariantizing the light-cone gauge, (\ref{twistor}) reduces to the relation between $\kappa-$symmetry fixed Type II GS superstring variables and the transverse physical NSR string variables $\varphi^{i}_\pm$. The equations of motion for the gauge fixed variables were obtained and solved. It was demonstrated that the gauge fixed Type II GS and NSR actions contain covariantized light-cone terms (\ref{gslightcone},\ref{lightcone}).
\section{Acknowledgements}
The author would like to thank A.A. Zheltukhin for the numerous valuable discussions, I.A.~Bandos and D.~Polyakov for the interest to the work and stimulating discussions, and the Abdus Salam ICTP, where the part of this work was done, for the warm hospitality. The work was supported by Ukrainian State Foundation for Fundamental Research.

\end{document}